\documentstyle[12pt]{article}   

\large
\newtheorem{definition}{Definition}[section]

\makeatletter
\@addtoreset{equation}{section}
\makeatother

\newcommand{\beq}{\begin{equation}}
\newcommand{\eeq}{\end{equation}}
\newcommand{\be}{\begin{equation}}
\newcommand{\ee}{\end{equation}}
\newcommand{\ba}{\begin{eqnarray}}
\newcommand{\ea}{\end{eqnarray}}

\newtheorem{Theorem}{Theorem}[section]

\newtheorem{Lemma}{Lemma}[section]

\def\Ac{A^\Co}
\def\Acb{\overline{A^\Co}}

\def\be{\begin{equation}}
\def\ee{\end{equation}}
\def\ba{\begin{eqnarray}}
\def\ea{\end{eqnarray}}

\def\agb{{\overline {{\cal A}/{\cal G}}}}

\def\Comp{{\mathchoice
{\setbox0=\hbox{$\displaystyle\rm C$}\hbox{\hbox to0pt
{\kern0.4\wd0\vrule height0.9\ht0\hss}\box0}}
{\setbox0=\hbox{$\textstyle\rm C$}\hbox{\hbox to0pt
{\kern0.4\wd0\vrule height0.9\ht0\hss}\box0}}
{\setbox0=\hbox{$\scriptstyle\rm C$}\hbox{\hbox to0pt
{\kern0.4\wd0\vrule height0.9\ht0\hss}\box0}}
{\setbox0=\hbox{$\scriptscriptstyle\rm C$}\hbox{\hbox to0pt
{\kern0.4\wd0\vrule height0.9\ht0\hss}\box0}}}}
\def\Co{{\mathchoice
{\setbox0=\hbox{$\displaystyle\rm C$}\hbox{\hbox to0pt
{\kern0.4\wd0\vrule height0.9\ht0\hss}\box0}}
{\setbox0=\hbox{$\textstyle\rm C$}\hbox{\hbox to0pt
{\kern0.4\wd0\vrule height0.9\ht0\hss}\box0}}
{\setbox0=\hbox{$\scriptstyle\rm C$}\hbox{\hbox to0pt
{\kern0.4\wd0\vrule height0.9\ht0\hss}\box0}}
{\setbox0=\hbox{$\scriptscriptstyle\rm C$}\hbox{\hbox to0pt
{\kern0.4\wd0\vrule height0.9\ht0\hss}\box0}}}}
\def\Rl{{\mathchoice
{\setbox0=\hbox{$\displaystyle\rm R$}\hbox{\hbox to0pt
{\kern0.4\wd0\vrule height0.9\ht0\hss}\box0}}
{\setbox0=\hbox{$\textstyle\rm R$}\hbox{\hbox to0pt
{\kern0.4\wd0\vrule height0.9\ht0\hss}\box0}}
{\setbox0=\hbox{$\scriptstyle\rm R$}\hbox{\hbox to0pt
{\kern0.4\wd0\vrule height0.9\ht0\hss}\box0}}
{\setbox0=\hbox{$\scriptscriptstyle\rm R$}\hbox{\hbox to0pt
{\kern0.4\wd0\vrule height0.9\ht0\hss}\box0}}}}

\def\agbc{{\overline { {\cal A}^\Comp/{\cal G}^\Comp}}}

\def\agbc{{\overline {{\cal A}^\Co/{\cal G}^\Co}}}

\title{An account of transforms on $\agb$}
\author{T. Thiemann\thanks{New address :
Physics Department, Harvard University, Cambridge, MA 02138, USA,
Internet : thiemann@math.harvard.edu}\\
Physics Department,\\
The Pennsylvania State University,\\
University Park, PA 16802, USA}
\date{{\small Preprint CGPG-95/11-3, Preprint HUTMP-95/B-347}}

\begin{document}

\maketitle                     

\begin{abstract}
In this article we summarize and describe the recently found transforms
for theories of connections modulo gauge transformations associated
with compact gauge groups. Specifically, we put into a coherent picture the
so-called loop transform, the inverse loop transform, the
coherent state transform and finally the Wick rotation
transform which is the appropriate transform that incorporates the correct
reality conditions of quantum gravity when formulated as a dynamical theory
of connections while preserving the simple algebraic form of the Hamiltonian
constraint. \end{abstract}

\section{Introduction}

Recently, there has been significant progress in the development of
calculus
on the space $\agb$ of connections modulo gauge transformations for
compact gauge groups \cite{1,2,3,4,5,6,7,8,9,10,11,12,13}. Roughly, we
can say that if we take the Abelian $C^*$ algebra generated by the Wilson
loop functions (that is, the traces of holonomies around loops for smooth
connections) and compute its Gel'fand spectrum then we obtain the space
$\agb$ of distributional connections which are equipped with the topology
defined by requiring that the Wilson loop functionals are continuous
\cite{1}.\\
This space is not some abstract construction lacking any geometric
interpretation, it can be very concretely characterized as the space of
homomorphisms from the group of piecewise analytic loops (the group structure
is obtained by identifying the inverse loop with the tracing of the
original loop in the opposite direction) to the gauge group \cite{2}.
It should be
said at this point that we are restricting ourselves to the analytic
category for the sake of simplicity although all the results can be
transferred to the smooth category \cite{14} subject to some fairly
technical modifications. \\
Since the Gel'fand spectrum is a compact Hausdorff space, measures thereon
are in one to one correspondence with positive linear functionals on the
space of continuous functionals on $\agb$. Beautiful,
diffeomorphism-invariant measures have been constructed \cite{2,3}.\\
The space $\agb$ can also be characterized as the projective limit of
a certain self-consistent family of measure spaces \cite{5,6}.\\
Differential geometry on $\agb$ can be constructed as well \cite{7,8,9}.
The framework can be applied to gauge theories \cite{10,11} and to
quantum gravity \cite{12,13} when formulated as a theory of connections
\cite{15}.\\
\\
In this article we focus on summarizing the theory underlying the
recently found transforms associated with $\agb$. All statements will be
given without proof, the details can be found in the original literature.

In section 2 we review the loop transform which was heuristically defined for
the first time in \cite{16} and later made precise in \cite{1}. This
transform maps us from functions of connections to functions of loops,
that is, (singular) knots.

Section 3 defines the inverse loop transform \cite{17}. We display here
the loop network version of that transform, however, there is also an
edge-network version of this theorem \cite{17} (compare also \cite{17a}
for a definition of spin-networks).

Section 4 reviews the coherent state transform \cite{13}.
This transform maps us from functions of real connections to functions of
complex connections.

Section 5 outlines the Wick rotation transform \cite{18} which does
not only map us to functions of {\em some} complex connection but to a
complex connection which corresponds to a {\em particular} complexification
of the real phase space. It is only with this transform at our disposal
that we can use all the techniques developed for compact gauge groups.
In particular, only then do we incorporate the correct reality conditions
while keeping the algebraic form of the Hamiltonian (constraint) simple.\\
An interesting modification of this transform is considered in \cite{18b}
which we outline briefly.\\
We display the explicit form of the transform for the case of canonical
quantum gravity.

\section{The loop transform}

We provide only the absolutely necessary information in order to fix
the notation. For further details see \cite{12} and references therein.\\
The space $\agb$ of generalized connections modulo gauge transformations
is the Gel'fand spectrum of the Abelian $C^\star$ algebra generated by the
Wilson-loop functionals for smooth connections, that is, traces of the
holonomy for piecewise analytic loops in the base manifold $\Sigma$. As such
it is a compact Hausdorff space and therefore measures on that space
are in one to one correspondence with positive linear functionals on
$C(\agb)$.\\
A certain natural measure $\mu_0$ will play a very crucial role in this
article so that we go now into more details :\\
In what follows, $\gamma\subset\Sigma$ will always denote a finite,
unoriented, closed (i.e. each vertex is at least 2-valent),
piecewise analytic graph, meaning that it is the union of a finite number
of analytic edges and vertices. Its
fundamental group $\pi_1(\gamma)$ is then finitely generated by some loops
$\beta_1(\gamma),..,\beta_{n(\gamma)}(\gamma)$ which we fix once and for
all together with some orientation and which are based at some arbitrary but
fixed basepoint $p\in\Sigma$, $n(\gamma)$ the number of generators of the
fundamental
group of $\gamma$. A function $f$ on $\agb$ is said to be cylindrical
with respect to a graph $\gamma$, $f\in\mbox{Cyl}_\gamma(\agb)$, if it is a
function only of the
finite set of arguments $p_\gamma(A):=(h_{\beta_1}(A),..,h_{\beta_n}(A))$
where $h_\alpha(A)$ is the holonomy of $A$ along the loop $\alpha$.
A measure $\mu$ is now specified by its finite joint distributions
$\mu_\gamma$ which are defined by
\be \label{2.1}
\int_\agb d\mu(A) f(A)=\int_{G^n} d\mu_\gamma(g_1,..,g_n)
f_\gamma(g_1,..,g_n)
\ee
where $f=f_\gamma\circ p_\gamma$ and $f_\gamma\;:G^n\to\Co$. In order that
this definition makes sense we have to make sure that if we write
$f=f_\gamma\circ p_\gamma=f_{\gamma'}\circ p_{\gamma'}$ in two
different ways as a cylindrical function where
$\gamma\subset\gamma'$ is a subgraph of $\gamma'$, then we
should have that the so-called consistency conditions
\be \label{2.2}
\int_{G^n}d\mu_\gamma f_\gamma=\int_{G^{n'}} d\mu_{\gamma'} f_{\gamma'}
\ee
are satisfied.\\
The natural measure $\mu_0$ is the induced Haar measure, meaning that
$d\mu_{0,\gamma}(g_1,..,g_n)=d\mu_H(g_1)..d\mu_H(g_n)$. One can check that
the consistency conditions are satisfied \cite{2} and that the so defined
cylindrical measure has a $\sigma$-additive extension $\mu_0$ on the
projective limit measurable space $\agb$ of the family of measurable spaces
$\agb_\gamma$ \cite{5}. The space $\agb_\gamma$
is defined to be the set of all conjugacy classes of homomorphisms from the
loop group restricted
to $\gamma$ into the gauge group while $\agb$ is the set of all
conjugacy classes of homomorphisms
from the loop group into $G$ (the conjugation corresponds to the gauge
transformations at the basepoint). Note that the semi-group of loops
with respect to compositions of loops can be given a group structure by
identifying paths that are traversed in the opposite direction with the
inverse of the original path.\\
\\
Let us now recall some basic facts from harmonic analysis on compact gauge
groups \cite{18a}.\\
Recall that every representation of a compact group
is equivalent to a unitary one, so that we may restrict ourselves to
unitary representations in the sequel. Also,
every such representation is completely reducible.
\begin{definition}
Let $\{\pi\}$ denote the set of all finite dimensional, non-equivalent,
unitary, irreducible representations of the compact gauge group $G$, let
$d_\pi$
be the dimension of $\pi$ and let $\mu_H$ be the normalized Haar measure
on $G$.\\
For any $f\in L_1(G,d\mu_H)$ define the Fourier transform of $f$ by
\be \label{3.1}
\hat{f}^{ij}_\pi:=\int_G d\mu_H(g) \sqrt{d_\pi}\bar{\pi}_{ij}(g)
f(g),\;i,j=1,..,d_\pi\;. \ee
\end{definition}
Note that this definition makes sense because the matrix elements of
$\pi(g)$ are bounded by 1.
\begin{definition}
The Fourier transform of a function is said to be $\ell_1$ or $\ell_2$
respectively iff
\be \label{3.2}
||\hat{f}||_1:=\sum_\pi\sum_{i,j=1}^{d_\pi} \sqrt{d_\pi}
|f^{ij}_\pi|<\infty\mbox{ or }||\hat{f}||_2:=\sum_\pi\sum_{i,j=1}^{d_\pi}
|f^{ij}_\pi|^2<\infty  \;. \ee
\end{definition}
The Fourier series associated with a function $f$ on $G$ such that
$\hat{f}\in\ell_1$ is given by
\be \label{3.3}
\tilde{f}(g):=\sum_\pi\sum_{i,j=1}^{d_\pi} \hat{f}^{ij}_\pi
\pi_{ij}(g)\sqrt{d_\pi}\;. \ee
The analogue of the Plancherel theorem for $\Rl^n$ is the Peter\&Weyl
theorem.
\begin{Theorem}[Peter\&Weyl]
1) The functions $g\to\sqrt{d_\pi}\pi_{ij}(g),\;i,j=1,..,d_\pi$ form a
complete and orthonormal system on $L_2(G,d\mu_H)$.\\
2) For any $f\in L_2(G,d\mu_H)$ it holds that $f=\tilde{f}$ in
the norm $||\;||_2$
and the Fourier transform is a unitary map $\wedge\;:\;L_2(G,d\mu_H)
\to\ell_2$.
\end{Theorem}
Next we introduce a new notion necessary to capture the gauge invariant
cylindrical functions.
\begin{definition}
i) A loop network is a triple $(\gamma,\vec{\pi},\pi)$ consisting of a
graph $\gamma$, a vector $\vec{\pi}=(\pi_1,..,\pi_{n(\gamma)})$ of
irreducible representations of $G$ and an irreducible representation
$\pi$ of $G$ which takes values in the set of irreducible
representations
of $G$ contained in the decomposition into irreducibles of the tensor
product $\otimes_{k=1}^n \pi_k$.\\
ii) A loop-network state is a map from $\agb$ into $\Co$ defined by
\be \label{3.12}
T_{\gamma,\vec{\pi},\pi}(A):=\mbox{\rm tr}[\otimes_{k=1}^{n(\gamma)}
\pi_k(h_{\beta_k(\gamma)}(A))\cdot c(\vec{\pi},\pi)]
\ee
where the matrix $c$ is defined by
\be \label{3.13}
c(\vec{\pi},\pi):=\sqrt{\frac{\prod_{k=1}^{n(\gamma)} d_{\pi_k}}{d_\pi}}
\pi(1)\;. \ee
\end{definition}
Loop network states satisfy the following important properties.
\begin{Lemma}
i) Given a graph $\gamma$, the set of all loop network states provides for
an orthonormal basis of
$L_2(\agb_\gamma,d\mu_{0,\gamma})=
L_2(\agb,d\mu_0)\cap\mbox{Cyl}_\gamma(\agb)$.\\
ii) Given a graph $\gamma'$, consider all its subgraphs $\gamma<\gamma'$.
Remove all the loop network states on $\gamma'$ which are pull-backs
of loop-network states on $\gamma$. The collection of all loop-network
states so obtained provides for an orthonormal basis of $L_2(\agb,d\mu_0)$.
\end{Lemma}
The proof of this lemma relies on the Peter\&Weyl theorem.
We are now in the position to define the Fourier transform of a measure on
$\agb$.
\begin{definition}
The loop transform (Fourier transform, characteristic functional) of a
measure $\mu$ on $\agb$ is defined by
\be \label{3.23}
\chi_\mu(\gamma,\vec{\pi},\pi):=<\bar{T}_{\gamma,\vec{\pi},\pi}>:=
\int_\agb d\mu(A) \bar{T}_{\gamma,\vec{\pi},\pi}(A)
\ee
\end{definition}
This definition differs from the one given in \cite{1,9,16}, however, both
definitions are equivalent in the sense that they allow for a reconstruction
of $\mu$ according to the Riesz-Markov theorem \cite{19}. Namely, the
former definition is based on the vacuum expectation value of products of
Wilson loop functionals, and according to \cite{1,20}, these functions are
an overcomplete set of functions on $\agb$ (that is, they are subject to
Mandelstam identities) so that we can reexpress them in terms of loop
networks and vice versa which are linearly independent.\\
Note that the loop transform of a measure on $\agb$ is the analog of the
Fourier transform of a measure on the quantum configuration space
of quantum scalar field theory \cite{10,21} and the Riesz-Markov theorem
substitutes the Bochner theorem \cite{22}.

\section{The inverse loop transform}

Now let be given a functional $\chi$ on loop-networks. Provided it is
positive (note that there are no Mandelstam relations between loop
network states any more and that the product of loop network states is a
linear combination of loop network states) we know by the Riesz-Markov
theorem that there is a measure $\mu$ whose Fourier transform is given
by $\chi$. This measure will be known if we know its finite joint
distributions $d\mu_\gamma$ (i.e. the loop transforms with respect to any
finite graph $\gamma$) which are automatically form a self-consistent system
of measures whose projective limit (known to exist) gives us back $\mu$.
We now wish to compute these joint distributions.\\
This task is not entirely trivial because of the following : first it is
not clear whether the finite joint distributions of a measure on $\agb$
are completely regular with respect to the product Haar measure. But even
if we assume it was, that is, a formula like $d\mu_\gamma=\rho_\gamma
d\mu_{0,\gamma}$ is true with $\rho_\gamma$ a positive $L_1$ function
(because the constant functions are integrable) then we still do not
know whether $\rho_\gamma\in
L_2(\agb_\gamma,d\mu_{0,\gamma})$. If that were true then we could
simply make
use of the fact that loop networks provide for an orthonormal basis of
$L_2(\agb,d\mu_0)$ to conclude theorem 3.3 directly from the gauge
invariant version of the Peter\&Weyl theorem. This is, however, not
necessarily the case.
\begin{Lemma}
If the Fourier transform of a (complex) regular Borel measure $\mu$ on a
compact gauge group $G$ is in $\ell_1$ then it is absolutely
continuous with respect to the Haar measure on $G$.
\end{Lemma}
The theorem can obviously extended to any finite number of variables.\\
Next we need to prove the analogue of the inverse Fourier
transform for compact groups \cite{17}.
This theorem answers the question whether a function which is
only $L_1$ can be represented by its Fourier transform.
\begin{Theorem}
Let $f\in L_1(g,d\mu_H)$ be such that also $\hat{f}\in\ell_1$. Then
$f(g)=\tilde{f}(g)$ in the sense of $||\;||_1$.
\end{Theorem}
So the logic is as follows : given a positive linear functional compute
its Fourier transform. If it is in $\ell_1$ then we can write the finite
joint distributions of the measure corresponding to $\chi$ as a positive
$L_1$ function times the product Haar measure. By the preceding theorem
we know that this function coincides in the $L_1$ sense with the associated
Fourier series.
\begin{Theorem}
Let $\chi$ be a positive linear functional on $C(\agb)$. Then $\chi$ is
the loop transform of a positive regular Borel measure $\mu$ on $\agb$. If
for a given graph $\gamma$ with n generators the sequence
$\{\chi(\gamma,\vec{\pi},\pi)\sqrt{d_\pi\prod_{k=1}^n d_{\pi_k}}\}$ is in
$\ell_1$
then the finite joint distributions of $\mu_\gamma$ are $\mu_{0,\gamma}$
a.e. given by \be \label{3.24}
\frac{d\mu_\gamma(A)}{d\mu_{0,\gamma}(A)}=
\sum_{\vec{\pi}}\sum_{\pi\in\otimes_{k=1}^n\pi_k} \chi(\gamma,\vec{\pi},\pi)
T_{\gamma,\vec{\pi},\pi}(A) \;.
\ee
\end{Theorem}
The proof is a straightforward application of the inverse Fourier
transform.\\
In order to determine whether a given function $\chi$ from (singular) knots
into the complex numbers
arises as the loop transform of a measure one has to check two things :\\
1) All the identities that are satisfied by products of traces of holonomies
of loops have to be satisfied Mandelstam identities \cite{19}.
Alternatively, it has to be true that $\chi$ can be written purely in
terms of loop-network states.\\
2) It is a positive linear functional on any cylindrical subspace of
$C(\agb)$.\\
An example of a (singular) knot function that satisfies these criteria is
of course the Fourier
transform of any $\sigma$-additive measure on $\agb$. Let us look
at the Fourier transform of the measure $\mu_0$ which is even diffeomorphism
invariant so that $\chi$ is a singular knot invariant :
\be \label{4.1}
\chi_{\mu_0}(\gamma,\vec{\pi},\pi)=\delta_{\vec{\pi},\vec{0}}\delta_{\pi,0}
\ee
where $0$ denotes the trivial representation. In other words,
$\chi$ is non-vanishing only on the trivial loop network $1$. Therefore we
find for the
finite joint distribution precisely $\rho_\gamma(g_1,..,g_n)=1$.

\section{The coherent state transform}

For applications in canonical quantum gravity, which when formulated as a
theory of connections \cite{15} leads naturally to a theory of
{\em complexified}
$SU(2)$ connections in order that the associated Hamiltonian constraint
takes a simple algebraic form, all what we have said so far is inapplicable
since it was fundamental to our approach that the gauge group was compact.
Therefore we are interested in a transform which maps us from a theory
of real connections to a theory of complex connections in such a way that
the physics is unchanged.\\
The mathematical input that led to the definition of this transform came
from the beautiful paper \cite{23} which generalizes the classical
Segal-Bargman transform (which is a unitary transform from the Schr\"odinger
representation $L_2(\Rl^n,d^nx)$ into the holomorphic representation
$L_2(\Co^n,d\nu(z,\bar{z}))\cap\mbox{Hol}(\Co^n)$ of
square-integrable
functions (with respect to a certain measure $\nu$) which are holomorphic)
to the case where copies of $\Rl$ are replaced by copies of a compact gauge
group $G$.\\
The details can be described by an analogy with the classical transform :\\
Suppose we have a Hilbert space ${\cal H}=L_2(\Rl,dx)$ and decide to
work in a representation ${\cal H}_\Co$ in which the holomorphic extension
$z=x_\Co$ of $x$ is a diagonal. We wish to do this in such a way that
the two Hilbert spaces are isometric as not to change the physics. The
way that this can be done is as follows :\\
There is the Laplacian $\Delta:=-\frac{1}{2}\partial_x^2$ on $\Rl$ and
associated with
it we construct the heat kernel $\hat{R}_t:=\exp(-t\Delta),\;t>0$. We now
just
define the coherent state transform to be kernel convolution followed by
analytic extension, that is
\be \label{5.1}
(\hat{C}_t\psi)(z):=<x|\hat{R}_t|\psi>_{|x\to z}:=[\int_\Rl dy \rho_t(x,y)
\psi(y)]_{x=z}
\ee
where $\rho_t(x,y)=\exp(-(x-y)^2/2t)/\sqrt{2\pi t}$ is the kernel of
$\hat{R}_t$ which is real analytic and therefore allows for a unique
analytic extension.\\
The measure $\nu_t$ which turns this transform into a unitary one is
constructed by taking the Laplacian on $\Co$, that is,
$\Delta_\Co=-\frac{1}{2}(\partial_z^2+\partial_{\bar{z}}^2)$, construct the
kernel
$\mu_t(z_1,z_2)=\mu_t(z_1-z_2)$ associated with it and average over the
real direction in $\Co$ with the Lebesgue measure $dx$ to produce
\be \label{5.2}
\nu_t(z,\bar{z}):=\int_\Rl \mu_t(z-x)
dx,\;d\nu_t(z,\bar{z}):=\nu_t(z,\bar{z})dz\wedge d\bar{z}\;.
\ee
The coherent state transform for any unimodular, compact gauge group is
now constructed simply by translating step by step all the structures
that we have defined so far into the terminology of group theory :\\
The Lebesgue measure $dx$ is the unique (up to a positive constant)
translation invariant measure on the real line. Likewise, the same is true
for the Haar measure $d\mu_H$ on $G$. Accordingly we replace $L_2(\Rl,dx)$
by ${\cal H}:=L_2(G,d\mu_H)$. Next we notice that the real line is the
additive group of real numbers and so the translation invariant vector
field $\partial_x$ is replaced by the left invariant vector field $X_i:=
\mbox{tr}(g\tau_i\partial_g)$ where translation now means group
multiplication and $\tau_i$ are the generators of the Lie algebra.
Obviously the Laplacian on $G$ is now defined to be $\Delta:=-\frac{1}{2}
\delta^{ij}X_i X_j$ and the associated heat kernel by
$\hat{R}_t:=\exp(-t\hat{\Delta})$. The analytic extension of the variable
$x$ is by going to $z=x+iy$. Notice that the additive group of real
numbers is its own Lie algebra generated by $1$ and that $\Co$ is
generated by $1,i$. Therefore in the case of G we take
the complexification of the Lie algebra generated by $\tau_j,i\tau_j$, i.e.
we go from $x^j \tau_j$ to
$(x^j+iy^j)\tau_j$ and exponentiate it to get the analytic extension
$G^\Co$ of $G$. Likewise we define the Laplacian on $G^\Co$ to be
$\Delta_\Co:=-\frac{1}{2}\delta^{ij}[X_i^\Co X_j^\Co+\bar{X}_i^\Co
\bar{X}_j^\Co]$ where $X_i^\Co$ is the analytic extension of
the left invariant vector fields $X_i$ on $G$ and the bar means
complex conjugation). We can now compute
$\rho_t(g,h)=\rho_t(g h^{-1})$ and $\mu_t(g_\Co,h_\Co)=\mu_t(g_\Co
h_\Co^{-1})$ as before and arrive at
\be \label{5.3}
\nu_t(g_\Co,\bar{g}_\Co):=\int_\Rl \mu_t(g_\Co g^{-1})
d\mu_H(g),\;d\nu_t:=\nu_t(g_\Co,\bar{g}_\Co)d\mu_H^\Co(g_\Co,\bar{g})
\ee
where, of course, $d\mu_H^\Co$ is the Haar measure on $\Co$.
This completes the construction of the coherent state transform for
one copy of $G$ \cite{13}.\\
We now wish to apply this machinery to our graph theoretic framework. Thus,
we are led to take a graph $\gamma$ equipped with $n$ generators $\beta_I$
of its fundamental group and we assign to it a Laplacian on $G^n$ defined by
\be \label{5.4}
\Delta_\gamma(A):=\sum_{I=1}^n l(\beta_n)\Delta_I
\ee
where $\Delta_I$ is the Laplacian associated with $g_I:=h_{\beta_I}(A)$,
the holonomy around $\beta_I$ as defined above and $l(\beta_I)$ is a function
from loops into the positive real numbers (so that the generator of the
transform is still a positive operator on $G^n$).\\
We now plug $\Delta_\gamma$ into the machinery explained above (extended
in an obvious way to $G^n$, that is, we do everything for each copy of
the group separately). In particular, we obtain heat kernel measures
$\rho_{t,\gamma}(g_1,..,g_n)$ and $\nu_{t,\gamma}(g_i^\Co,\bar{g}_i^\Co)$,
however, in order that these measures define cylindrical measures the
function $l$ cannot be entirely arbitrary. We need to check the consistency
conditions that arise from the fact that we may write a cylindrical function
on different graphs. As one can easily see, if a function $f$ is cylindrical
with respect to a graph $\gamma$ then it is also cylindrical with respect
to any bigger graph $\gamma'>\gamma$, i.e. $f=p_\gamma^\ast f_\gamma
=p_{\gamma'}^\ast f_{\gamma'}$ and
it is easy to see that $\gamma'$
can be obtained from $\gamma$ by a finite number of steps of the following
type :\\ a) we just add one more generator $\beta_{n+1}$ independent of
$\beta_1,..,\beta_n$,\\
b) one of the generators of $\gamma$, $\beta_1$ say, is a composition of
generators of $\gamma'$, so a formula like
$\beta_1=\beta_i'\circ\beta_j'$ holds and\\
c) one generator just appears inverted, $\beta_1=(\beta_1')^{-1}$.\\
Now the requirement $\int d\rho_\gamma f_\gamma=\int d\rho_{\gamma'}
f_{\gamma'}$ leads to restrictions on $l$ as follows :\\
Requirement a) does not lead to a restriction since the heat kernel measures
are non-interacting and normalized but b) leads to $l(\alpha\circ\beta)
=l(\alpha)+l(\beta)$ and c) to $l(\alpha^{-1})=l(\alpha)$.\\
This completes the construction of the coherent state transform.\\
However, our framework is unsatisfactory for two reasons :\\
a) Recall that the purpose of the coherent state transform is to map us
from a
theory of real connections to a theory of complex connections. Since two
connections, real or complex, have the same transfromation properties
under diffeomorphisms, our transform should better be
generated by a diffeomorphism covariant operator. While $\Delta_I$ is
covariantly defined since it depends only on the holonomy of the
connection around $\beta_I$, the function $l$ clearly
breaks diffeomorphism covariance as it is independent of the connection,
it is just kinematically defined, and not diffeomorphism invariant.
So at best we can expect to work in a diffeomorphism-gauge fixed context.\\
b) Actually in \cite{13} also a diffeomorphism-covariant coherent state
transform
is constructed based on measures introduced by Baez, however, those measures
are not faithful and thus do not serve to provide inner products (they
provide sesquilinear forms which are not positive definite) so that they
cannot be used for our purposes.
b) The second disadvantage is that the only reason why to go from real
to complex connections, at least in the context of quantum gravity, was to
simplify the Hamiltonian constraint which is quite messy if we start, for
instance, with the spin-connection representation \cite{24}.
However, the present transform $\hat{R}_t$ when viewed as a transform
on ${\cal H}:=L_2(\agb,d\mu_0)$ does not map us from the spin
connection $\Gamma_a^i$ to the Ashtekar connection $\Gamma_a^i-iK_a^i$
\cite{15} since it commutes with $\Gamma_a^i$ and so does not lead to
a simplification of the Hamiltonian constraint at all. So the virtue of
the complex representation is lost. Moreover, the spin-connection can be
shown to be a bad coordinate for the triads \cite{18}

\section{The Wick rotation transform}

The Wick rotation transform is geared at fixing both problems mentioned
in the last paragraph in one stroke. In fact, it turns out that the
coherent state
transform can be viewed from a broader perspective as the unique solution
of how to identify the particular complexification of a real variable
with its analytic extension.

In the next subsection we will give an outline of the algorithm of
incorporating the correct
reality conditions into the quantum theory for a general theory. The
procedure naturally leads to a generalized coherent state transform for whose
underlying measure $\nu_t$, analogous to the ones discussed in the previous
section, we will give
a general formula. We comment on the available quantization strategies
which automatically incorporate the correct reality conditions while
keeping the constraints simple. Finally, we apply the algorithm to
canonical quantum gravity.

\subsection{The Complexifier and the Wick rotation}

Suppose we are given some real phase space
$\Gamma$ (finite or infinite dimensional) coordinatized by a canonical
pair $(A,P)$ (we suppress all labels like indices or coordinates) where
we would like to think of $A$ as the configuration variable and $P$ as its
conjugate momentum. Let
the Hamiltonian (constraint) on $\Gamma$ be given by a function $H(A,P)$
which has a quite complicated algebraic form and suppose that it turns
out that it can be written in polynomial form if we write it in terms of a
certain complex {\em canonical} pair $(A^\Co,P_\Co):=W^{-1}(A,P)$, that
is, the function $H_\Co:=H\circ W$ is polynomial in $(A^\Co,P_\Co)$ (the
reason why we begin with the inverse of the invertible map $W$ will become
clear in a moment). We will not be talking about kinematical constraints
like Gauss and diffeomorphism constraint etc. which take simple algebraic
forms in any kind of variables.\\
The requirement that the complex pair $A^\Co,P_\Co$ is still canonical is
fundamental to our approach and should be stressed at this point. It
should also be stressed from the outset that we are {\em not}
complexifying the phase space, we just happen to find it to convenient to
coordinatize it by a complex valued set of functions. The reality
conditions on these functions are encoded in the map $W$. \\
Of course, the theory will be much easier to solve (for instance
computing the spectrum (kernel) of the Hamiltonian (constraint)
operator) in a holomorphic representation ${\cal H}_\Co$ in which
the operator corresponding to
$A_\Co$ is diagonal rather than in the real representation $\cal H$
in which the operator corresponding to $A$ is diagonal.
According to the canonical commutation relations and in order
to keep the Hamiltonian (constraint) as simple as possible, we are
naturally led to represent the operators on $\cal H$ corresponding
to the canonical pair $(A,P)$ by $(\hat{A}\psi)(A)=A\psi(A),
(\hat{P}\psi)(A)=-i\hbar\delta\psi(A)/\delta A$. Note that then in
order to meet the adjointness condition that $(\hat{A},\hat{P})$ be
self-adjoint
on $\cal H$, we are forced to choose ${\cal H}=L_2({\cal C},d\mu_0)$ where
$\cal C$ is the quantum configuration space of the underlying theory and
$d\mu_0$ is the unique (up to a positive constant) uniform
measure on $\cal C$, that is, the Haar measure.\\
In order to avoid confusion we introduce the following notation
throughout this section :\\
Denote by $\hat{K}:\;{\cal H}\to{\cal H}_\Co$ and $\hat{K}^{-1}$
the operators of analytic continuation and restriction to real arguments
respectively. The operators corresponding to $A^\Co,P_\Co$ can be
represented on the two distinct Hilbert spaces ${\cal H}$ and
${\cal H}_\Co$. On $\cal H$ we just define them by some ordering
of the function $W^{-1}$, namely
$(\hat{A}^\Co,\hat{P}_\Co):=W^{-1}(\hat{A},\hat{P})$. On
${\cal H}_\Co$, the fact that $A^\Co,P_\Co$ enjoy canonical brackets
allows us to define their operator versions simply by (and this is why it
is important to have a {\em canonical} complex pair)
$(\hat{A}',\hat{P}')=(\hat{K}\hat{A}\hat{K}^{-1},
\hat{K}\hat{P}\hat{K}^{-1})$, i.e. they are just the
analytic extension of $\hat{A},\hat{P}$, that is,
$(\hat{A}'\psi)(A^\Co)=A^\Co\psi(A^\Co),(\hat{P}'\psi)(A^\Co)=
-i\hbar\delta\psi(A^\Co)/\delta A^\Co$.\\
But how do we know that the operators
$\hat{A}^\Co,\hat{P}_\Co$ on $\cal H$ and $\hat{A}',\hat{P}'$
on ${\cal H}_\Co$ are the quantum analogues of the same classical
functions $A^\Co,P_\Co$ on $\Gamma$ ?
To show that there is essentially only one answer to this
question is the first main result of this section.\\
Namely, when can two operators defined
on different Hilbert spaces be identified as different representations
of the same abstract operator ? They can be identified iff their matrix
elements coincide. Due to the canonical commutation relations we have
to identify in particular also the matrix elements of the identity
operator, that is, scalar products between elements of the Hilbert spaces.
The only way that this is possible is to achieve that
the Hilbert spaces are related by a unitary transformation and that the
two operators under question are just images of each other under this
transformation.\\
In order to find such a unitary transformation we have to relate the
two sets of operators $\hat{A}^\Co,\hat{P}_\Co$ and $\hat{A}',\hat{P}'$
via their common origin of definition, namely through the set
$\hat{A},\hat{P}$.\\
The first hint of how to do this comes from the observation that both pairs
$(A,P)$ and
$(A^\Co,P_\Co)$ enjoy the same canonical commutation relations, i.e. they
are related by a {\em canonical complexification}.
Therefore the map $W$ must be a complex symplectomorphism, that is, an
automorphism $(W\cdot O)(A,P)=(O\circ W)(A,P)$
of the Poisson algebra of (not necessarily real-valued) functions $O$ over
$\Gamma$ (in particular $A,P$)
which preserves the algebra structure (linear and symplectic structure)
but, of course, not the reality structure. Let $iC$ be the infinitesimal
generator of this automorphism (we do not require that $C$ be a
bounded, positive or at least real functional), that is,
\be \label{7.1}
O_\Co=W^{-1}\cdot O =\sum_{n=0}^\infty
\frac{(-i)^n}{n!}\{O,C\}_{(n)}
\ee
where, as usually, the multiple Poisson bracket is iteratively defined by
$\{O,C\}_{(0)}=f,\;\{O,C\}_{(n+1)}=\{\{O,C\}_{(n)},O\}$. It
follows immediately from (\ref{7.1}) that the
reality conditions for $O$ are given by (the bar denotes complex
conjugation)
\be \label{7.3}
\bar{O}_\Co =\sum_{n=0}^\infty \frac{i^n}{n!}
\{O_\Co,C+\bar{C}\}_{(n)}\;.
\ee
We are now going to assume that the functional $C$ has a well-defined
quantum analogue, that is, the {\em Complexifier} $\hat{C}$ will be a
(not necessarily bounded,
not necessarily positive, not necessarily self-adjoint) densely
defined operator on $\cal H$. Further, we would like to
take equations (\ref{7.1}),(\ref{7.3}) over to quantum theory, that is, we
replace Poisson brackets by commutators times $1/i\hbar$ and we replace
complex conjugation by the adjoint operation on ${\cal H}=L_2({\cal
C},d\mu_0)$.\\
So let $\hat{O}=O(\hat{A},\hat{P})$ be any operator on $\cal H$ where
$O$ is an analytical function.
Using the operator
identity $e^{-\hat{f}}\hat{g}e^{\hat{f}}=\sum_{n=0}^\infty\frac{1}{n!}
[\hat{g},\hat{f}]_{(n)}$ and defining on $\cal H$
\be \label{7.4} \hat{W}_t:=\exp(-t\hat{C}) \ee
we find that on $\cal H$ we may {\em define} the quantum version of
(\ref{7.1}) by
\be \label{7.5}
\hat{O}_\Co:=O(\hat{A}^\Co,\hat{P}_\Co)=\hat{W}_{-t}\hat{O}\hat{W}_t
\ee
with $t=1/\hbar$. That is, the generator $\hat{C}$ provides
for a natural ordering of the function $W^{-1}(A,P)$.\\
$\hat{W}_t$ is called the {\em Wick rotator} (due to
its role in quantum gravity). \\
We are now in the position to make the identification of the two sets
$\hat{A}',\hat{P}'$ and $\hat{A}^\Co,\hat{P}_\Co$ precise, by
writing the operators on
${\cal H}_\Co$ in terms of the operators on $\cal H$. We have
\be \label{6.1}
(\hat{A}',\hat{P}')= (\hat{K}\hat{A}\hat{K}^{-1},
\hat{K}\hat{P}\hat{K}^{-1})=
(\hat{U}_t\hat{A}^\Co\hat{U}_t^{-1},\hat{U}_t\hat{P}_\Co\hat{U}_t^{-1})
\ee
where we have defined
\be \label{6.2}
\hat{U}_t:=\hat{K}\hat{W}_t\;.
\ee
So if we could achieve that $\hat{U}_t$ is a {\em unitary} operator from
$\cal H$ to ${\cal H}_\Co$ then our identification would be complete ! \\
The way to do that is, of course, by constructing an appropriate  measure
$\nu_t$ on the complexified quantum configuration space ${\cal C}_\Co$
thereby fixing the Hilbert space ${\cal H}_\Co$ to be the space of
square-integrable functions on ${\cal C}_\Co$ with respect to $d\nu_t$
which are also holomorphic, that is,
${\cal H}_\Co:=L_2({\cal C}_\Co,d\nu_t)\cap\mbox{Hol}({\cal C}_\Co)$ .
We see that $\hat{U}_t$ precisely coincides with the coherent
state transform of the previous section in case that $\hat{C}$ is the
Laplacian on the group so that we call it the {\em generalized coherent state
transform} and, due to its role in quantum gravity, we will also refer
to it as the {\em Wick (rotation) transform}. In other words, the
Wick transform can be
viewed as the {\em unique} answer to our question. Any other unitary
transformation $\hat{u}_t$ between the Hilbert spaces necessarily
corresponds to a different complexification
$\hat{w}_t=\hat{K}^{-1}\hat{u}_t$ of the classical phase space in which the
Hamiltonian (constraint) takes a more complicated appearance.
Note that any real canonical transformation corresponds to a unitary
transformation in quantum theory, so the coherent state transform can
also be characterized as the ``unitarization" of the complex canonical
transformation that we are dealing with.\\
Another characterization of the coherent state transform $\hat{U}$ follows
from the simple formula $\hat{K}=\hat{U}_t\hat{W}_t^{-1}$ : it is the unique
solution to the problem of how to identify
analytic extension with the particular choice of complex
coordinates $A^\Co,P_\Co)$ on the real phase space $\Gamma$ as defined by
$\hat{W}^{-1}$.\\
It should be stressed at this point that if $\hat{C}$ is not a positive
self-adjoint operator (so that $\hat{W}_t$ is not a
well-defined operator on $\cal H$ for $t>0$ which explains why we
started with $W^{-1}$ rather than $W$), we
will assume that $\hat{U}_t$ for positive $t$ can still be densely
defined (note that $\hat{U}_{-t}\not=\hat{U}_t^{-1}$ due to the analytic
continuation involved), that is, there is a dense subset $\Phi$ of $\cal
H$ so that
the analytic continuation of the elements of its image $\hat{W}_t\Phi$
under $\hat{W}_t$ are elements of a dense subset of ${\cal H}_\Co$.
{\em We do not assume that $\hat{U}_t$ itself can be densely defined on
$\Phi$ as an operator on $\cal H$} ! Intuitively what happens here is
that while $\hat{W}_t\phi$ may not be normalizable with respect to $\mu_0$
for any $\phi\in\Phi$, its analytic continuation will be normalizable with
respect to $\nu_t$ by construction since $\hat{U}_t$ is unitary and thus
bounded, just because the measure $\nu_t$ falls off much stronger at
infinity than $\mu_0$. So we see that going to the complex representation
could be forced on us. This is a second characterization of the coherent
state transform : not only is it a unique way to identify a particular
complexification with analytic continuation, it also provides us with the
necessary flexibility to choose a better behaved measure $\nu_t$ which
enables us to work in a representation in which $\hat{W}_t$
or, rather, $\hat{W}_t'$ is well-defined which is important because only then
do we quantize the original theory defined by $({\cal H},\hat{H})$.\\
As a bonus, our adjointness relations are trivially incorporated !
Namely, because any operator $\hat{O}_\Co$ on $\cal H$ written in terms of
$\hat{A}^\Co,
\hat{P}_\Co$ is defined by $\hat{W}_t^{-1}\hat{O}\hat{W}_t$ where $\hat{O}$
is written in terms of $\hat{A},\hat{P}$ and because $\hat{O}_\Co$ is
identified on ${\cal H}_\Co$ with
$\hat{O}'=\hat{K}\hat{O}\hat{K}^{-1}=\hat{U}_t\hat{O}_\Co\hat{U}_t^{-1}$ we
find due to unitarity that
\be \label{7.5b}
(\hat{O}')^\dagger=\hat{U}_t\hat{O}_\Co^\dagger
\hat{U}_t^{-1}
\ee
where the adjoints involved on the left and right hand side of
this equation are taken on ${\cal H}_\Co$ and $\cal H$ respectively.
Therefore,
$(\hat{O}')^\dagger$ is identified with $\hat{O}_\Co^\dagger$ as
required.\\
Note that the adjoint of $\hat{O}_\Co$ on $\cal H$
\be \label{7.5a}
\hat{O}_\Co^\dagger =
[\hat{W}_t^\dagger\hat{W}_t]\hat{O}_\Co[\hat{W}_t^\dagger\hat{W}_t]^{-1}
\ee
follows unambiguously from the known
adjoints of $\hat{A},\hat{P}$ and coincides to zeroth order in $\hbar$ with
the complex conjugate of its classical analogue (\ref{7.3}) and therefore
can be seen to be one particular operator-ordered version of
the adjointness relations that follow from the requirement that the
classical reality conditions (\ref{7.3}) should be promoted to
adjointness-relations in the quantum theory.\\
Interestingly, $\hat{W}_t^\Co=\hat{W}_t$ on $\cal H$ corresponding to the
fact that classically the complexifier is unchanged if we replace
$A,P$ by $A^\Co,P_\Co$.\\
Finally we see that in extending the algebraic programme \cite{12}
from a real representation to the holomporphic representation
of the Weyl relations we only have one additional input, everything else
follows from the
machinery explained below and can be summarized as follows :\\
Input A : define an automorphism $W$ (preferrably such that the
constraints simplify).\\
Task A : determine the infinitesimal generator $C$ of $W$.\\
Input B : define a real $^*$ representation $\cal H$.\\
Task B : determine a holomorphic representation ${\cal H}_\Co$ so that
$\hat{U}=\hat{K}\hat{W}$ is unitary.\\
Note that input B is also part of the programme if one were dealing
only with the real representation so that input A is the only additional
one. Task A is necessary if we
want to express a given $W$ in terms of the phase space variables
which is unavoidable in order to define $\hat{W}$.\\
In the next subsection we display a standard solution to Task B so that Task
A is the only non-trivial problem left .\\
Remark :\\
Note that the existence of $W$ does not imply that classical solutions
are mapped
into solutions ! That is, assume that we have found a physical solution
$H(A_0,P_0)=E=const.$, then in general
$H_\Co(A_0,P_0)=H(W(A_0,P_0))\not=const.$ (this has nothing to do
with the fact that $W$ does not preserve the reality conditions,
rather it follows from the fact that $\{H,C\}\not=0$ by
construction since $W$ is supposed to turn the complicated
algebraic form of $H$ into a simpler one). However, it will turn out that
the quantum analogue of
$W$ maps generalized eigenfunctions into generalized eigenfunctions !

\subsection{Isometry}

The construction of $\nu_t$ for a general theory differs considerably from
the techniques applied in \cite{13} and the previous section
which turn out to be insufficient in dealing with an operator
$\hat{C}$ which is neither positive nor self-adjoint. For instance, in
applications to quantum gravity we need to apply the more general theory
given below. We will only display the result here, the details can be
found in \cite{18}.

Isometry means that for any $\psi,\xi\in{\cal H}$ we have
\be \label{7.8}
\int_\agb d\mu_0(A)\overline{\psi}(A)\xi(A)=\int_\agbc d\nu_t(\Ac,\Acb)
\overline{[\hat{U}_t\psi](\Ac)}[\hat{U}_t\xi](\Ac) \;.
\ee
Denote by $\mu_0^\Co(\Ac)$ the holomorphic extension of
$\mu_0$ and by $\bar{\mu}_0^\Co(\Acb)$ its anti-holomorphic extension which
due to the positivity of
$\mu_0$ are just complex conjugates of each other. We now make the
ansatz
\be \label{7.9}
d\nu_t(\Ac,\Acb)=d\mu_0^\Co(\Ac)\otimes
d\bar{\mu}_0^\Co(\Acb)\nu_t(\Ac,\Acb)\;, \ee
where $\nu_t$ is a distribution. Then we find that
equation (\ref{7.8}) can be solved by requiring
\be \label{7.11}
\nu_t(\Ac,\Acb):=\left({\overline{\hat{W}_{-t}^\dagger}}\right)'
\overline{\left({\overline{\hat{W}_{-t}^\dagger}}\right)'}\delta(\Ac,\Acb)
\ee
where the distribution involved in (\ref{7.11}) is defined by
\be \label{7.12}
\int_\agbc d\mu_0^\Co(\Ac)d\bar{\mu}_0^\Co(\Acb) f(\Ac,\Acb)
\delta(\Ac,\Acb)=\int_\agb d\mu_0(A) f(A,A) \;.
\ee
Here the adjoints are taken with respect to $\mu_0$, the overline denotes
complex conjugation of the whole expression of the operator (in particular
$\Ac\to \Acb,\delta/\delta\Ac\to
\delta/\delta\Acb$) and {\em not} anti-analytic extension and finally
the prime means analytic extension
$A\to A^\Co,\delta/\delta A\to\delta/\delta A^\Co$ as before.
Whenever $(\ref{7.11})$ exists and the steps to obtain this formula can
be justified we have proved existence of an isometricity inducing
positive measure on $\agbc$ by explicit construction. The rigorous proof
for this \cite{25} is by proving existence of (\ref{7.11}) on
cylindrical subspaces, so strictly speaking $d\nu_t$ is only a cylindrical
measure. The measure is self-consistently defined because the operator
$\hat{C}$ is. \\
Note that the proof is immediate in the
case in which $\hat{C}$ is a positive and self-adjoint and therefore can be
viewed by itself
as an interesting extension of \cite{13}. In particular it coincides
with the method introduced by Hall \cite{23} in those cases when $\hat{C}$ is
the Laplacian on $G$ but our technique allows for a more straightforward
computation of $\nu_t$.

\subsection{Quantization}

We are now equipped with two Hilbert spaces ${\cal H}:=L_2(\agb,d\mu_0)$
and ${\cal H}_\Co:=L_2(\agbc,d\nu_t)\cap\mbox{Hol}(\agbc)$ which are
isometric and faithfully
implement the adjointness relations among the basic variables. $\cal H$
will be called the real representation and ${\cal H}_\Co$ the holomorphic
or complex representation.\\
The last step in the algebraic quantization programme is to solve the theory,
that is, to find the spectrum of the Hamiltonian (or the kernel of the
Hamiltonian constraint) and observables, that is, operators that commute
with the physical constraint operators. In more concrete terms it means the
following \cite{12} :\\
Let $\hat{H}_\Co:=H_\Co(\hat{A},\hat{P})$ be a convenient ordering of
$H_\Co(A,P)$ such that its adjoint on $\cal H$ corresponds to the
complex conjugate of its classical analogue (that is, write $H_\Co=a+ib$
where $a,b$ are real and order $a,b$ symmetrically) and
let $\hat{H}'=H_\Co(\hat{A}',\hat{P}')$ be its analytic extension.
Choose a topological vector space $\Phi(\Phi_\Co)$
and denote by $\Phi'(\Phi_\Co')$ its topological dual. These two spaces are
paired by means of the measure $\mu_0(\nu_t)$, for instance
$f[\phi]:=\int_\agb d\mu_0(A)\bar{f}[A]\phi[A]$. $\Phi(\Phi_\Co)$ is by
construction dense in its Hilbert space completion ${\cal H}({\cal H}_\Co)$.
We will be looking for generalized eigenvectors
$f_\lambda(f_\lambda^\Co)$ \cite{23a}, that is,
elements of $\Phi'(\Phi_\Co')$ with the property that there exists a complex
number $\lambda$ such that $f_\lambda[\hat{H}_\Co\phi]=\lambda
f_\lambda[\phi](f_\lambda^\Co[\hat{H}'\phi]=\lambda
f_\lambda^\Co[\phi])$ for any $\phi\in\Phi(\Phi_\Co)$.\\
Given this general setting we have at least two strategies at our disposal :\\
Strategy I) :\\
We start working on ${\cal H}_\Co$. This means that we would
try to find a convenient ordering of the operator
$\hat{H}':=H_\Co(\hat{A}',\hat{P}')$. The Hamiltonian (constraint) on
$\cal H$ now is {\em defined} to be the image under the {\em inverse}
coherent state transform $\hat{H}:=\hat{U}_t^{-1}\hat{H}'\hat{U}_t$
which to zeroth order in $\hbar$ coincides with
one ordering of $H(A,P)$ but in general will involve an infinite power
series in $\hbar$. That is, we have made use of the freedom that we
always have in defining the quantum analog of a classical function, namely
to add arbitrary terms which are of higher order in $\hbar$.\\
Of course, since the constraint is simple on ${\cal H}_\Co$ we solve it
in this representation as well as the problem of finding observables.
After that we can go back to $\cal H$ which is technically easier to handle
and compute spectra of the observables found and so on, thus making use
of the powerful calculus on $\agb$ that has already been developed in
\cite{2,3,4,5,6,7,8,9,10,11,12}. This calculus can in particular be used
to find a regularization in which $\hat{C},\hat{H}_\Co=H_\Co(\hat{A}^\Co,
\hat{P}_\Co)$ are self-adjoint
on ${\cal H}$ if they classically correspond to real functions because
it then follows that also their images under $\hat{W}_{-t}$, that is,
$\hat{C},\hat{H}$, are self-adjoint on $\cal H$ and then, by unitarity,
the same holds for $\hat{C}',\hat{H}'$ on ${\cal H}_\Co$.\\
In this way, ${\cal H}_\Co$ mainly arises as an intermediate step to solve
the spectral problem.\\
Strategy II) :\\
The following strategy is suggested by Abhay Ashtekar \cite{18b}
in the restricted context of quantum gravity. Namely, just stick with the
real representation all the time !
The general idea idea of working only with real connections goes back to
\cite{15} and was revived by Barbero in \cite{27}.
This strategy now seems feasible because now we have a {\em key} new
ingredient at our disposal -- the Wick transform (compare next
subsection) -- which enables the real representation to simplify
both, the reality conditions and the constraints.\\
Here we consider this strategy in the general case.
That means,
we look for a convenient ordering of $\hat{H}_\Co:=H_\Co(\hat{A},\hat{P})$,
then the physical Hamiltonian (constraint) is {\em defined} by
$\hat{W}_{-t} \hat{H}_\Co \hat{W}_t$ and agrees to zeroth order in
$\hbar$
with some ordering of $H(\hat{A},\hat{P})$. The advantage of this approach
is obvious : we never need to construct the measure $\nu_t$ which is only
cylindrical so far while the measure $\mu_0$ is known to be
$\sigma-$additive. For instance in the case of quantum gravity,
although we continue to work with a connection
dynamics formulations, the {\em complex} connection,
drops out of the game altogether ! All the results in
\cite{2,3,4,5,6,7,8,9,10,11,12} are immediately available while in
strategy II) one could do so only after having solved the spectral problem
(constraint).\\
Why then, would we ever try to quantize along the lines of strategy I) ?
This is because there can be in general obstructions to find
the physical spectrum or kernel directly from
$\hat{H}_\Co$. This is also the reason why we extended the programme as to
construct the coherent state transform.
Such obstructions can arise as follows : $H_\Co(A,P)$ will
be in general neither positive nor real valued even if $H(A,P)$ is, in
which case it is questionable
whether one is in principle able to find the correct spectrum from
the former Hamiltonian (constraint). This is so because both
$\hat{W}_t$ and $\hat{U}_t$ preserve the spectrum wherever they are
defined, meaning that if we fail to have coinciding spectra for
$\hat{H},\hat{H}_\Co$ then $\hat{W}_t$ is ill-defined as a map on $\cal
H$ or on the dense subset $\Phi$ while $\hat{U}_t$ is always well-defined on
$\Phi$ (by construction) as an operator between $\cal H$ and ${\cal
H}_\Co$ since it is unitary. \\
A fortunate case is when the topological vector space $\Phi$ is preserved by
$\hat{W}_t$ : then generalized eigenvectors $f_\lambda$ of $\hat{H}_\Co$ are
mapped (as elements of the topological dual space $\Phi'$)
into generalized eigenvectors $\hat{W}_t^\dagger f_\lambda$ of
$\hat{H}$ with
the same eigenvalue. The proof is easy : we have for any $\phi\in\Phi$ :\\
$\hat{W}_t^\dagger f_\lambda[\hat{H}\phi]=f_\lambda[\hat{W}_t\hat{H}\phi]
=f_\lambda[\hat{H}_\Co\hat{W}_t\phi]=\lambda \hat{W}_t^\dagger
f_\lambda[\phi]$ as claimed. Note that it was crucial in this argument
that $\hat{W}_t\phi\in\Phi$.\\
There are indications \cite{25} that we are lucky in the case of quantum
gravity.\\
Another way out might be the following : if $\hat{W}_t$ does not preserve
$\Phi$ then we may just reduce the size of $\Phi$, thereby enlarging $\Phi'$,
as to turn all the formal solutions to the constraints into rigorously
defined generalized eigenvectors.\\
A, minor, disadvantage of strategy II) is as follows :
while we can find physical observables $\hat{O}$ by looking for operators
$\hat{O}_\Co$ that commute with $\hat{H}_\Co$ and then map them
according to $\hat{O}:=\hat{W}_{-t}\hat{O}_\Co\hat{W}_t$, since
$\hat{W}_t$ is not unitary on $\cal H$ we need to compute the expectation
values $<\psi,\hat{W}_{-2t}\hat{O}_\Co\xi>$ rather than just
$<\psi,\hat{O}_\Co\xi>$. Via strategy I) we could compute everything
either on $\cal H$ or ${\cal H}_\Co$, whatever is more convenient, because
$\hat{U}_t$ is unitary and so does not change expectation values.\\
\\
To summarize : \\
If we proceed along strategy I) then we quantize the same
physical Hamiltonian (constraint)
$\hat{H}':=H_\Co(\hat{A}',\hat{P}')$ and
$\hat{U}_t^{-1}\hat{H}\hat{U}_t$ in two different but
unitarily equivalent representations ${\cal H}_\Co$ and $\cal H$. The more
convenient representation is
${\cal H}_\Co$ because the Hamiltonian (constraint) adopts a simple form.
This procedure is guaranteed to lead to the correct physical spectrum of
oservables while it is technically more difficult to carry out since we
are asked to construct the measure $\nu_t$.

If we proceed along strategy II) then we quantize the two distinct
Hamiltonians $\hat{H}_\Co:=H_\Co(\hat{A},\hat{P})$ and $\hat{H}:=
\hat{W}_{-t}\hat{H}_\Co\hat{W}_t$ (of which the latter is the physical
one) in the same representation $\cal H$. While this procedure is
technically easier to perform, as explained above, its validity
depends on the strong condition that $\hat{W}_t$ can be densely defined on
$\cal H$ which is often not the case !

\subsection{A transform for quantum gravity}

Denote by $q_{ab},K_{ab}$ the induced metric and extrinsic curvature of a
spacelike hypersurface $\Sigma$, introduce a triad $e_a^i$ which is an
$SU(2)$ valued one-form by $q_{ab}=e_a^i e_b^j\delta_{ij}$ and denote
by $e^a_i$ its inverse. Then we introduce the canonical pair of Palatini
gravity by $(K_a^i:=K_{ab} e^b_i,P^a_i:=1/\kappa \sqrt{\det(q)}e^a_i)$ where
$\kappa$ is Newton's constant.\\
We will now employ our algorithm to find the coherent state transform for
quantum gravity.\\
The important observation due to Ashtekar \cite{15} is that if we write
the theory in terms of the complex canonical pair $A_a^{\Co j}:=\Gamma_a^j
-iK_a^j,P^a_{\Co j}:=i P^a_j$ where $\Gamma$ is the spin-connection
associated with $P$, then
the Hamiltonian constraint adopts the very simple polynomial form
$H_\Co(A_\Co,P^\Co)=-\mbox{tr}(F_{ab}^\Co[P^a_\Co,P^b_\Co])$ where $F^\Co$
is the curvature of $A^\Co$. The importance of this observation is that
$(A^\Co,P_\Co)$ is a {\em canonical} pair which relies on the discovery that
the spin connection is integrable with generating functional
$F=\int_\Sigma d^3x \Gamma_a^i P^a_i$. Ashtekar and later Barbero \cite{27}
also considered the real {\em canonical} pair
$(A_a^j=\Gamma_a^j+K_a^j,P^a_j)$
in which, however, the Hamiltonian takes a much more complicated
non-polynomial, algebraic appearence which becomes
polynomial only after multiplying by a power of
$\det(P^a_i)$ which turns the constraint into an unmanagable form.
After neglecting a term proportional to the Gauss constraint this
Hamiltonian can be written as
$H_\Co(A^\Co,P_\Co)\equiv H(A,P)=\mbox{tr}(\{F_{ab}-2R_{ab}\}[P^a,P^b])$ in
which $F,R$ are respectively the curvatures of $A,P$.\\
The real and complex canonical pairs are related by a chain of three canonical
transformations $(A=\Gamma+K,P)\to(K,P)\to(-iK,iP)\to(A^\Co=\Gamma-iK,P_\Co
=iP)$ of which the first and the last have as its infinitesimal generator
the functional $-F$ and $iF$ respectively. The new result to is that
we are able to derive the infinitesimal generator of the middle
simplectomorphism $(K,P)\to(-iK,iP)$ which is a {\em phase space Wick
rotation} !\\
This generator is uniqely given by (compare \cite{18} for a systematic
derivation)
\be \label{8.1}
C:=\frac{\pi}{2}\int_\Sigma d^3x K_a^i P^a_i
\ee
which is easily recognized as ($\pi/2\kappa$ times) the integral over the
densitized trace of the extrinsic curvature of $\Sigma$.\\
The key observation in proving this is that the Poisson bracket of
$C$ with the spin-connection $\Gamma_a^i$ vanishes.
The elegant way of seeing it is by noticing that $\tilde{C}$ generates
constant scale transformations and remembering that $\Gamma_a^i$ is a
homogenous rational function in $P_a^i$ and its spatial derivatives of
degree zero. Summarizing, the Ashtekar variables
$(A_a^{\Co,j}=\Gamma_a^j
-i K_a^j,P^a_{\Co,j}=i P^a_j)$ are the result of a Wick rotation
generated by $C$ in the sense of (\ref{7.1}), namely
\ba \label{8.3}
A_a^{\Co,j}(x) &=&\sum_{n=0}^\infty \frac{(-i)^n}{n!}\{A_a^j(x),C\}_{(n)}
=\Gamma_a^i(x)+[\sum_{n=0}^\infty\frac{(-i\pi/2)^n}{2}]K_a^i(x)
\nonumber\\
& = & \Gamma_a^i(x)+e^{-i\pi/2}K_a^i(x)
\ea
and similarily for $P^a_i$.\\
The unphysical Hamiltonian constraint
$H_\Co(A,P)=-\mbox{tr}(F_{ab}[P^a,P^b])$
is up to the negative sign just the Hamitonian constraint of the formal
Hamiltonian formulation of Euclidean gravity (it is easy to see that
our Wick rotated Lorentzian action equals that of Euclidean gravity
if we replace the lapse by its negative and the shift and Lagrange multiplier
of the Gauss constraint by $-i$ times themselves), however it should be
stressed that what we are doing here is not the quantization of Euclidean
gravity : there is no analytic continuation of the time coordinate involved
for which there is no natural choice anyway.\\
What is important is that $\hat{C},\hat{H}_\Co$ on $\cal H$ can be
chosen to
be self-adjoint operators and regularized with the techniques already
available in the literature because it is a classically real expressions.
It is a pecularity of the gravitational Hamiltonian that $H$ and
$H_\Co$ are both real.\\
According to our general programme we have completed Input A, namely we
have chosen the Ashtekar complexification, and task A, we have found the
complexifier $C$. The next Input B is to choose a $^\ast$ representation.
We choose
${\cal H}:=L_2(\agb,d\mu_0)$, that is ${\cal C}=\agb$ is the quantum
configuration space of real distributional connections introduced in
section 2 and $d\mu_0$ is precisely the induced Haar measure
on $\agb$ also introduced in section 2 which is uniquely selected
by the requirement that loop and
strip operators are essentially self-adjoint \cite{12}.\\
The task B left to do is to construct the measure $\nu_t$ for the
Hilbert space
${\cal H}_\Co:=L_2(\agbc,d\nu_t)\cap\mbox{Hol}(\agbc)$, the set of square
integrable functions of complexified connections which are holomorphic. Here
$\agbc$ is the quantum configuration space of complexified connections
modulo gauge transformations \cite{13}.
This task requires to find an appropriate regularization of the
operator version of $C$.
This seems to be a hopeless thing to do because when written in terms of
$A,P$ it involves the spin-connection which is a non-polynomial
expression. However,
using that for $\sqrt{|\det(E)|}$ already a well-defined regularization
exists \cite{28}, in \cite{25} a regularization is considered
which defines a self-adjoint operator, well-defined on
diffeomorphism-invariant
states (in the same sense as in \cite{29}) and which
leaves every cylindrical subspace {\em separately} invariant. This operator
is not positive so the latter property is quite important if one wants to
exponentiate it (for instance the regularization used in \cite{29}
does not have this property).\\
We can then define the (Wick or) ``generalized coherent state transform"
associated with the ``heat
kernel" $\hat{W}_t$ for the operator $\hat{C}$ to be the following map
\be \label{7.6} \hat{U}_t\; : {\cal H}\to {\cal H}_\Co, \;
\hat{U}_t[f](A_\Co):=<A_\Co,\hat{W}_t f>:=<A,\hat{W}_t f>_{|A\to A^\Co} \ee
which on functions cylindrical with respect to a graph generated
by n loops $\beta_I$ reduces to (because $\hat{C}$ leaves that subspace
invariant)
\be \label{7.7} \hat{U}_{t,\gamma}[f_\gamma](g^\Co_1,..,g^\Co_n):=
\int_{G^n}d\mu_{0,\gamma}(g_1,..,g_n)\rho_{t,\gamma}(g^\Co_1,..,
g^\Co_n\;;\;g_1,..,g_n) f_\gamma(g_1,..,g_n) \ee
where $g_I:=h_{\beta_I}(A),g^\Co_I:=h_{\beta_I}(A^\Co)$ are the holonomies
along the loop $\beta_I$. Here $\rho_{t,\gamma}(g_I\;;\; h_I):=<g_1,..,g_n,
\exp(-t\hat{C}_\gamma) h_1,..,h_n>$ is the kernel of $\hat{W}_t$ and
$\hat{C}_\gamma$ is the projection of $\hat{C}$ to the given cylindrical
subspace of $L_2(\agb,d\mu_0)$. The kernel, if it
exists, is real analytic on $G^n$ and therefore has a unique analytic
extension.\\
Note that the transform is consistently defined on cylindrical subspaces
of the Hilbert space because its generator $\hat{C}$ acts primarily on the
connection and does not care how we write a given cylindrical function
on graphs that are related to each other by inclusion.\\
\\
\\
\\
{\large Acknowledgements}\\
\\
I thank the organizers of the Sintra workshop ``Recent
mathematical developments in classical and quantum gravity'', July 26-28th
1995, for inviting me.
I am grateful for many important insights obtained in the course of
discussions with Abhay Ashtekar, Jurek Lewandowski, Donald
Marolf and Jos\'e Mour\~ao. \\
The author was supported in part by the NSF Grant PHY93-96246 and the Eberly
research fund of The Pennsylvania State University.

\end{document}